# Simulation analysis of rectangular dielectric-loaded traveling wave amplifiers for THz-sources

Changbiao Wang
ShangGang Group, 470 Prospect Street, Apt. 72, New Haven, CT 0651, USA

Non-linear simulation results for a 220-GHz rectangular dielectric-loaded traveling wave amplifier are presented. Simulations are used to check a linear theory that is developed by phenomenological introduction of an effective dielectric parameter for electron beam channel, and it is found that the RF power gains from Pierce three-wave theory and particle simulations are in reasonable agreement. It is shown that the RF power gain during initial beam-wave interaction is positive; the falling on the initial RF power profile, which has been thought to be the RF power transferred to the beam for bunching buildup (negative gain effect), is probably resulting from numerical errors. Beam-wave interaction mechanism is analyzed by examining the evolution of beam bunching centers. Influences of various parameters on amplifier performance are examined, and transverse space-charge effect is analyzed. A symmetric excitation scheme for RF couplers is proposed, and RF field jumps on the common intersection line of vacuum, dielectric, and metal wall, which is found in RF simulations, are explained theoretically.

## I. INTRODUCTION

Terahertz (100 GHz ~ 30 THz) technology has extensive applications and aroused increasing interest.[1,2] In addition to bound atomic or molecular state-based lasers, bremsstrahlung-based free-electron devices have been used for THz sources, such as free-electron lasers,[3] coherent synchrotron light sources[4,5] and radiators,[6,7] backward wave oscillators,[8] and Smith-Purcell radiators.[9] Gyrotrons are high power RF sources since their interaction structures can be several times of operating wavelength. Recently, a 3.5-kV low-voltage CW gyrotron was demonstrated experimentally, with an output of 12 W at 233 GHz.[10] As the development of the technology of micromachining and carbon nanotube cold cathodes,[11] a novel miniature vacuum electron device, 1.2-THz nanoklystron has been under design and fabrication.[12]

A lot of attention has been paid to investigation on dielectric-loaded traveling wave tubes (TWT), mostly with a cylindrical configuration,[13,14,15,16] while the rectangular configuration catching less interest. A rectangular structure may have some potential advantages; for example, RF input/output couplers may have a larger bandwidth because a symmetric excitation can be made to match the RF fields of operating mode, and dielectric slabs are easier to fabricate, especially for short wavelength applications.

In this paper, simulation analysis of a 220-GHz rectangular dielectric-loaded TWT amplifier is presented. This amplifier powered by an electron beam of a few tens of keV and mA operates at LSM mode. The LSM mode is more effective in the TWT interaction, although LSE mode, as proposed by Ling and Liu for a slow wave free-electron laser,[17] has no electric component at the direction normal to the dielectric slabs and may reduce deflection of electrons to hit dielectric walls. The beam stability can be obtained through an applied focusing magnetic field.

In the study, a simplified simulation method is adopted, where Lagrangian perspective for particle dynamics is used to count the action of RF fields on electrons, while the reaction of electrons on RF fields is considered through energy conservation law, that is, $P_b + P_{rf} + P_{loss}$ = constant, where $P_b$ is the beam power, $P_{rf}$ is the flowing RF power, and $P_{loss}$ is the total loss power caused by conducting walls and dielectrics. It is assumed in this method that only the forward traveling wave of design $LSM_{21}$ mode[18] is excited, and the transverse space-charge effect can be considered through a simplified analytic model. This method is essentially the same as the one used to study cyclotron autoresonance acceleration.[19]

TWT linear theory can provide an intuitive physical explanation, while non-linear theory can reveal physical results that linear theory does not include, and they are complementary tools. In this study, simulations are used to check a linear theory that is developed by phenomenological introduction of an effective dielectric parameter. Beam-wave interaction mechanism is analyzed by examining the evolution of bunching centers. Two kinds of RF power absorption in the initial beam-wave interaction process are illustrated: one is from Pierce three-wave theory that predicts a negative gain for a *loss-free* circuit; the other is from particle simulations for a *lossy* circuit. A numerical RF power absorption in particle simulations for a *loss-free* circuit is identified, which has been thought to be the RF power transferred to the beam for bunching buildup, leading to a negative gain.[16,20] Influences of various parameters on TWT performance are examined, and transverse space-charge effect is analyzed. A proof-of-principle simulation design of RF couplers with symmetric excitations for rectangular dielectric-loaded TWTs is presented.

The cross section for the longitudinally uniform structure used in simulations is shown in Fig. 1(a) and the field profiles for the LSM$_{21}$ mode is shown in Fig. 1(b), with structure dimensions tabulated in Table 1. The dielectric constant is taken to be 12; for example, silicon material has such a dielectric property[21] and it is easy to make micromachining. Figure 2 shows the dispersion curves for this structure. It is seen from Fig. 2 that the phase velocity of LSM$_{21}$ mode at 220 GHz exactly corresponds to a 30-keV electron energy. Increasing the ratio of the height 2$d$ to the width 2$b$ may increase mode separations and improve the uniformity of $E_z$-field in the vacuum beam channel, but it will weaken RF fields for a given flowing power; increasing dielectric constant may reduce beam voltage, but the RF fields in the beam channel will get closer to dielectric surfaces.

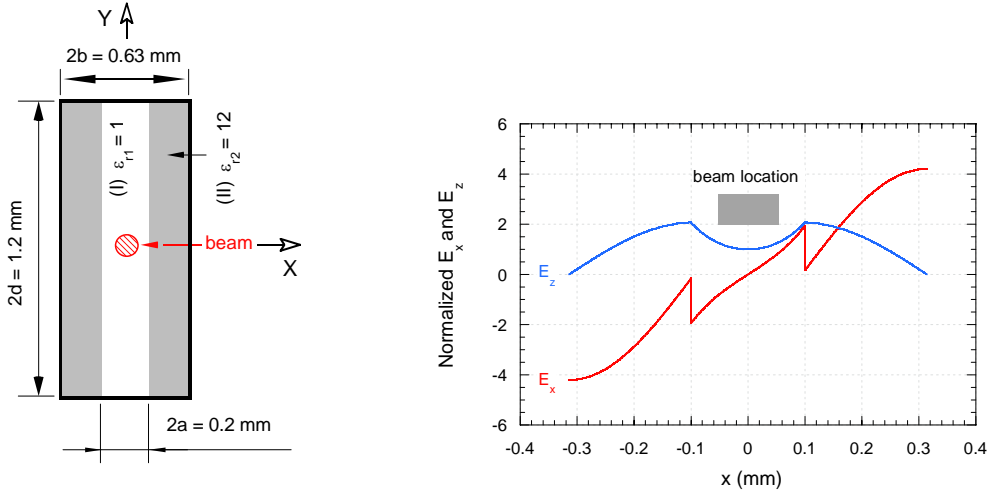

(a). Cross section of interaction region.  (b). Normalized electric field profiles for LSM$_{21}$ mode.

Fig. 1. Configuration for a 30-keV, 220-GHz TWT structure and RF electric field distributions from analytic theory. The interaction region is a piece of symmetric three-zone rectangular dielectric-loaded waveguide. $E_x$ and $E_z$ are mirror-symmetric with respect to y-z plane; $E_{xmax}$ = 38.6 kV/m and $E_z$(center) = 9.2 kV/m for a 1-W RF flowing power.

Table 1. TWT parameters and structure dimensions (mm) used in simulations.

| nominal frequency (GHz) | 220 | dielectric constant | 12 |
| --- | --- | --- | --- |
| waveguide height 2$d$ | 1.2 | beam channel width 2$a$ | 0.2 |
| waveguide width 2$b$ | 0.628 | dielectric thickness $b - a$ | 0.214 |

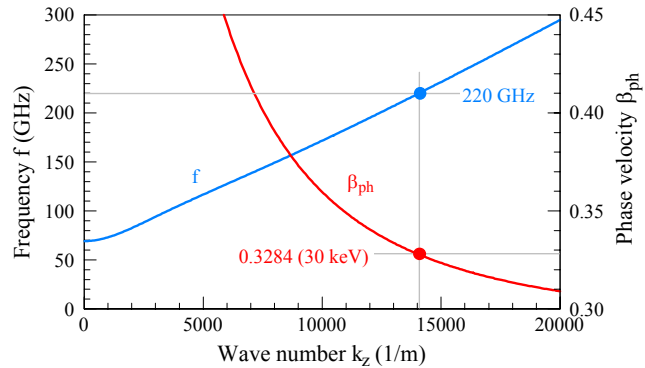

Fig. 2. Dispersion curves: frequency $f$ and normalized phase velocity $\beta_{ph}$ vs axial wave number $k_z$ for LSM$_{21}$ operating mode. It is seen that at 220 GHz ($k_z$ = 14042/m and $\lambda g = 2\pi/k_z$ = 0.4475 mm), a 30-keV electron beam synchronizes with the mode.

## II. COMPARISON BETWEEN LINEAR AND SIMULATION RESULTS

Various linear theories have been developed to analyze beam-wave interaction in dielectric-loaded structures.[22,23] In this section, an alternative linear theory is presented, that is developed by introduction of an effective dielectric parameter for the beam channel, and then particle simulations for a cold beam is used to check the linear theory.

The beam-wave interaction equation in this linear theory is obtained based on standard linear assumptions for a cold electron beam coupling with a single $mn$-th waveguide mode in a longitudinally translation-invariant dielectric-loaded waveguide structure. By ignoring the transverse perturbations and multimode coupling effects, the equation can be written as

$$\left(k_z^2 - \frac{\omega^2}{c^2}F_{diel} + \widetilde{k}_{\perp mn}^2\right)\left(k_z - \frac{\omega}{v_0}\right)^2 = -\frac{\omega_p^2}{v_0^2}\left(\frac{\omega^2}{c^2}F_{diel} - k_z^2\right)F_{beam}, \qquad (1)$$

where

$$\widetilde{k}_{\perp mn}^2 = \frac{\omega_{mn}^2}{c^2}F_{diel} - k_{zmn}^2, \quad \text{and} \quad \omega_p = \left(\frac{e\rho_0}{\varepsilon^{(b)}m_0\gamma_0^3}\right)^{1/2}. \qquad (2)$$

In the above, the eigen frequency $\omega_{mn}$ and eigen wave number $k_{zmn}$ are related through $\omega_{mn} = k_{zmn}v_0$, with $v_0$ the beam velocity; $\gamma_0 = (1-\beta_0^2)^{-1/2}$ is the beam relativistic energy factor, with $\beta_0$ the beam velocity normalized to the vacuum light speed $c$; $\omega_p$ is the generalized plasma frequency; $\varepsilon^{(b)}$ is the dielectric permittivity in the beam channel; $e$ and $m_0$ are the electron's charge and rest mass, and $\rho_0$ is the uniform equilibrium charge density; the dielectric filling factor $F_{diel}$ and the beam filling factor $F_{beam}$ are, respectively, given by

$$F_{diel} = \Pi^{-1}\iint_S \mu_r\varepsilon_r\left(\varepsilon E_{zmn}E_{zmn}^* + \mu H_{zmn}H_{zmn}^*\right)dS, \qquad (3)$$

$$F_{beam} = \Pi^{-1}\iint_{S_b} \varepsilon E_{zmn}E_{zmn}^* dS, \qquad (4)$$

where $\Pi = \iint_S (\varepsilon E_{zmn}E_{zmn}^* + \mu H_{zmn}H_{zmn}^*)dS$ with $E_{zmn}$ and $H_{zmn}$ the eigen mode axial electric and magnetic fields, and $(\varepsilon_r) \varepsilon$ and $(\mu_r) \mu$ the (relative) permittivity and permeability, $S$ denotes the waveguide cross section, and $S_b$ denotes the beam cross section.

If the structure is fully filled with only one dielectric and the beam overlaps the dielectric, then $F_{diel} = \varepsilon_r\mu_r$ and $F_{beam} = 1$ for TM modes ($H_z = 0$), and the beam-wave interaction Eq. (1) is reduced to the one given by Schächter.[24] Therefore, Eq. (1) can be taken to be a generalization from that equation just by modifying a key factor $[(\omega^2/c^2)\mu_r^{(b)}\varepsilon_r^{(b)}-k_z^2]$ to be $[(\omega^2/c^2)F_{diel}-k_z^2]$, where the *product* of $\varepsilon_r^{(b)}$ and $\mu_r^{(b)}$ in the beam channel is replaced by $F_{diel}$, as seen in the right-hand side of Eq. (1). The reason for this modification can be explained as follows.

Unlike the generalized plasma frequency $\omega_p$ that is only decided by the equilibrium beam parameters and local dielectric property, the propagation property of radiation waves in the vacuum beam channel of a multi-zone dielectric-loaded structure is affected by the surrounding loaded dielectric, and the channel behaves like a dielectric that has an "effective" dielectric parameter. It is seen from Eq. (3) that the dielectric filling factor $F_{diel}$ is such an effective parameter. As shown below, the physical results obtained based on this phenomenological modification are reasonably self-consistent, and the linear growth rate and power gain are comparable with the ones from particle simulations.

One might conjecture that directly using $\mu_0\varepsilon_0$ to replace $\mu^{(b)}\varepsilon^{(b)}$ of the vacuum beam channel, a growing wave solution also can be obtained because the cubic roots of (+1) are involved in analytically solving Eq. (1). Unfortunately, that growing wave solution is contrary to energy conservation law although its growth rate is not unacceptable compared with the one from particle simulations.

It is a widely accepted practice in the community that real frequency $\omega$ and complex axial wave number $k_z$ are assumed in the TWT linear theory when calculating beam-wave interaction equation.[24,25] The roots of Eq. (1) for $k_z$ at $\omega = \omega_{mn} = k_{zmn}v_0$ (eigen wave frequency) can be approximately estimated, given by

$$k_z = \begin{cases} (k_{zmn} + \Delta/2) + i\Delta\sqrt{3}/2, & \text{(growing wave)} \\ (k_{zmn} + \Delta/2) - i\Delta\sqrt{3}/2, & \text{(decaying wave)}, \\ (k_{zmn} - \Delta), & \text{(normal wave)} \end{cases} \qquad (5)$$

where

$$\Delta \equiv \left( \frac{\omega_{p0}^2}{\omega_{mn}^2} \frac{(\beta_0^2 F_{diel} - 1) F_{beam}}{2\gamma_0^3 \varepsilon_r^{(b)}} \right)^{1/3} k_{zmn}, \tag{6}$$

with $\omega_{p0} = (e\rho_0/\varepsilon_0 m_0)^{1/2}$ the classic plasma frequency. From Eq. (5), we find that $(k_z - k_{zmn})$ is equal to $\Delta$ multiplied by the cubic roots of (-1). For one dielectric-filled structures with a beam overlapping the dielectric, $\varepsilon_r^{(b)} = \varepsilon_r$, $F_{diel} = \varepsilon_r \mu_r$, and $F_{beam} = 1$ for TM mode; thus Eq. (6) is greatly simplified. For multi-zone structures which have a vacuum beam channel, $\varepsilon_r^{(b)} = 1$ holds but computations are needed to evaluate the dielectric filling factor $F_{diel}$ and the beam filling factor $F_{beam}$.

It is seen from Eq. (5) that the growing and decaying waves are slow waves compared with the eigen wave of which the phase velocity is equal to the beam velocity $v_0$, while the normal wave is a fast wave. This property is consistent with the one given by Andrews and Brau.[23] For a 30-keV, 25-mA line beam interacting with $LSM_{21}$ mode ($F_{diel} = 11.7$ and $k_{z21} = 14042$/m), the numerical results directly from Eq. (1) are $|Im(k_z)| = 40.2$/m (growth rate), $Re(k_z) = 14065$/m for growing and decaying waves, and $k_z = 13994$/m for normal wave; they are in good agreement with the ones estimated from Eq. (5) where $\Delta = 46.8$/m given by Eq. (6).

Figure 3 shows the RF gain profile based on Pierce three-wave linear theory,[26] compared with the ones from particle simulations using 0.1-mW, 1-mW, and 10-mW input RF powers. The profile from three-wave theory is reasonably close to the ones from simulations before they reach saturations. According to the linear theory, the gain profile has nothing to do with the input RF power. From Fig. 3 it is seen that the gain profiles from simulations also share this property from beginnings to before saturations. Simulations show that the 30.2-keV result better fits the three-wave result, compared with the 30.15-keV and 30.25-keV results.

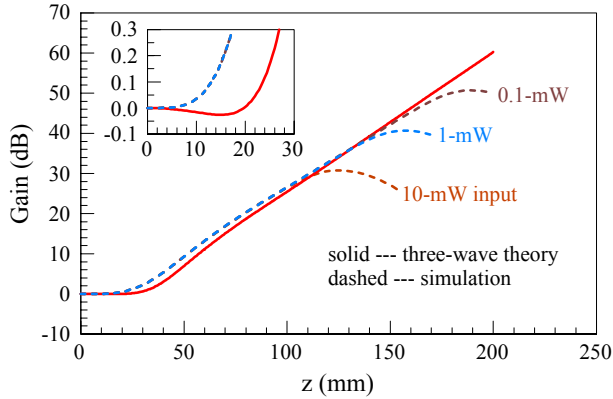

Fig. 3. Gain profile from Pierce three-wave theory (solid) compared with the ones from particle simulations (dashed) for 0.1-mW, 1-mW, and 10-mW input RF powers. Three-wave theory: 30-keV, 25-mA cold line beam; simulation: 30.2-keV, 25-mA cold line beam. The gains from simulations are positive in the initial beam-wave interaction process while the gain from Pierce three-wave theory is negative.

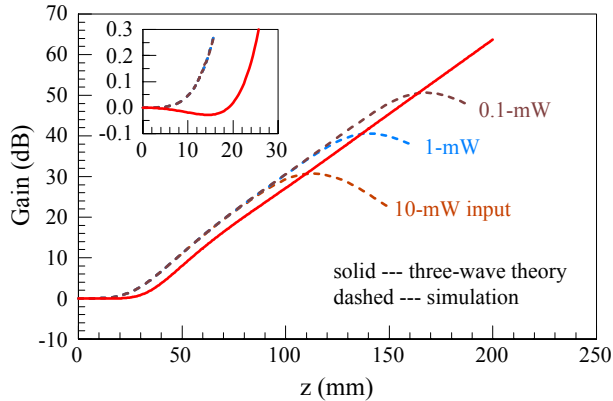

Fig. 4. Gain profiles from Pierce three-wave linear theory (solid) and particle simulations (dashed) for a 0.1mm×0.1mm beam focused by a 4-kG magnetic field.

It should be noted from Fig. 3 that the power gain from particle simulations for a loss-free circuit is positive in the initial beam-wave interaction process while the power gain from Pierce three-wave theory is negative, which is one of the distinctive differences between particle simulations and Pierce three-wave theory. Another well-known difference is that the linear theory cannot describe saturation of beam-wave interaction.

It is interesting to point out that a proper choice of numerical methods in simulations is significant. For example, if a direct method for taking an average is employed in particle simulations, considerable numerical errors may result in an "RF power absorption" appearing during the initial beam-wave interaction, which was explained to be the negative gain effect in the initial beam bunching buildup process in some literature (see Appendix).

For a finite size beam, the results from three-wave theory and simulations have a little larger deviations but comparable, as shown in Fig. 4. The cold beam used in Fig. 4 has a transverse dimension of 0.1mm×0.1mm. To focus the beam in simulations, a 4-kG magnetic field is applied. It is seen that the deviations can be up to 3.4 dB, but the gain profiles from simulations all intersect the one from three-wave theory before saturations.

The simplest case for analysis of Čerenkov radiations is the one dielectric-filled structure with a fully-filled beam although it is less practical. It is well recognized physically that the electron beam generates Čerenkov radiation waves in such a way that all the waves have the same phase velocity equal to the beam velocity, that is, $v_{ph} = \omega/k_z = v_0$ or $(k_z - \omega/v_0) = 0$,[18,27] while the waveguide structure requires any electromagnetic waves to satisfy the waveguide dispersion equation $\omega^2 \mu\varepsilon - k_\perp^2 - k_z^2 = 0$ with $k_\perp$ the constant transverse wave number decided by boundary conditions; as a result, only those radiation waves that simultaneously meet the two physical conditions can exist, leading to a discrete radiation spectrum. Accordingly, from the viewpoint of coupling-wave theory, the Čerenkov TWT mechanism can be explained to be the radiation wave coupling with the waveguide eigen modes. Because of the coupling the wave number is modified, as shown in Eq. (5).

The expressions of the perturbed charge and current densities used in deriving beam-wave interaction Eq. (1) are given by

$$\begin{pmatrix} \rho^{(1)} \\ J_z^{(1)} \end{pmatrix} = -\begin{pmatrix} k_z \\ \omega \end{pmatrix} \frac{i\varepsilon_0 \omega_{p0}^2 E_z}{\gamma_0^3 (\omega - k_z v_0)^2} , \qquad (7)$$

where the axial RF electric field is given by $E_z = E_{zmn} \exp[i(\omega t - k_z z)]$ with $E_{zmn}$ the eigen mode field.[18] The beam-wave interaction power density $\mathrm{Re}(J_z^{(1)} E_z^*)$ may be larger or less than zero, depending on growing wave or decaying wave. From Eqs. (5) and (7) we have $\mathrm{Re}(J_z^{(1)} E_z^*) < 0$ at $\omega = \omega_{mn}$ for growing wave because there is a $5\pi/6$-phase difference between $J_z^{(1)}$ and $E_z$, while $\mathrm{Re}(J_z^{(1)} E_z^*) > 0$ for decaying wave ($\pi/6$-phase difference). That means that the growing RF wave does negative work on the beam while the decaying RF wave does positive work, which is consistent with energy conservation law. Of course, the normal wave does no work because the phase difference between $J_z^{(1)}$ and $E_z$ is $\pi/2$. From above analysis, we can understand why the solution related to the cubic roots of (+1) will violate energy conservation law.[22]

From Eq. (5), it is seen that the growing and decaying waves have the same modified phase velocity, and the beam velocity is larger than the wave phase velocity no matter for growing wave or for decaying wave; however, simulations show that the beam may support a decaying wave only when the beam velocity is less than the phase velocity.

## III. DESCRIPTION OF TWT MECHANISM

In non-linear particle simulations, beam-wave interaction process in the TWT amplifiers can be roughly divided into three parts: buildup of beam bunching (low-gain lethargy region[16,28]), wave amplification, and saturation. In the bunching-buildup process, particle phase density gets modulated to form bunching centers in decelerating phases. As bunching centers drift towards the maximum deceleration field, the beam gives up more energy while the bunching is further strengthened so that the RF wave gets effectively amplified, until the beam phase is saturated with significant part of particles getting out of deceleration fields. In this section, the interaction mechanism is examined by simulation analysis of a cold line beam.

To better understand the TWT process, first let us take a qualitative look at the evolution of a particle's phase under the influence of RF field. For a particle on the *z*-axis, the phase can be written as

$$\omega t - k_z z \approx \omega t_0 - (k_z/\beta_0) \int_0^z [(\beta_0 - \beta_{ph}) + \delta\beta] dz , \qquad (8)$$

where $\beta_{ph}$ is the phase velocity normalized to the vacuum light speed, $\delta\beta = \beta - \beta_0$, with $\beta$ the normalized particle velocity and $\beta_0 = \beta$ at $z = 0$, and $\omega t_0$ is the particle injection phase. In obtaining Eq. (8), $|\delta\beta/\beta_0| \ll 1$ and $\beta_0 \approx \beta_{ph}$ are used. Suppose that all the particles are uniformly injected along the *z*-axis with the same initial velocity $\beta_0 > \beta_{ph}$, so that $(\beta_0 - \beta_{ph}) + \delta\beta = (\beta - \beta_{ph}) > 0$ holds before losing substantial energy. It is seen from Eq. (8) that the phase ($\omega t - k_z z$) reduces as the particle moves. If a particle is accelerated, then $\delta\beta$ (>0) increases, leading to ($\omega t - k_z z$) reducing faster; if a particle is decelerated, then $\delta\beta$ (<0) decreases, leading to ($\omega t - k_z z$) reducing slower. The effective field component responsible for the beam bunching is the axial component, given by $E_z = E_0 \sin(\omega t - k_z z)$ on the z-axis with $E_0 > 0$. Therefore, the axial electric field $E_z > 0$ when $0 < (\omega t - k_z z) < \pi$, where the electrons are decelerated and their phases reduce slower; while the electrons in $\pi < (\omega t - k_z z) < 2\pi$ are accelerated and their phases reduce faster, as if the decelerated particles were waiting for the accelerated particles to come together, resulting in phase bunching with $\omega t_0 = \pi$ the initial bunching center. Since the equilibrium beam velocity $\beta_0$ is larger than the phase velocity $\beta_{ph}$, the initial bunching center drifts into the deceleration field as particles move.

From above qualitative analysis we know that the physical basis for beam bunching comes from the fact that the phase of a particle being accelerated reduces faster while the phase of a particle being decelerated reduces slower. Below, a quantitative examination of the beam bunching is given for a cold line beam on the z-axis in the 220-GHz rectangular dielectric-loaded TWT structure.

The bunching mechanism in slow wave traveling wave tubes was well described qualitatively in a simplified way where the observer is assumed to sit in the RF wave frame.[29] There is an alternative way that is often used to quantitatively describe the mechanism by particle simulations, where one judges a bunching process by observing the change of a particle distribution on the energy-phase plane during beam-wave interaction.[16] Usually it is not easy to directly identify where the bunching center is and how the center shifts. Here we provide a more intuitive way to show the evolution of particle phase density or beam bunching centers during the interaction.

In the particle simulations, all 320 beam particles are initially uniformly distributed in one RF cycle from $\omega t_0 = 0$ to $2\pi$. Figure 5 shows the dependence of RF power and gain on axial interaction distance for the line beam mentioned above, without any conducting and dielectric loss included. There is an approximate linear growth regime around z = 50.2 mm with a growth rate of 46.0/m. As indicated previously, the RF power gain in the low gain region (z < 50.2 mm) is always larger than zero, different from the one given by Schächter where the power gain is less than zero.[16]

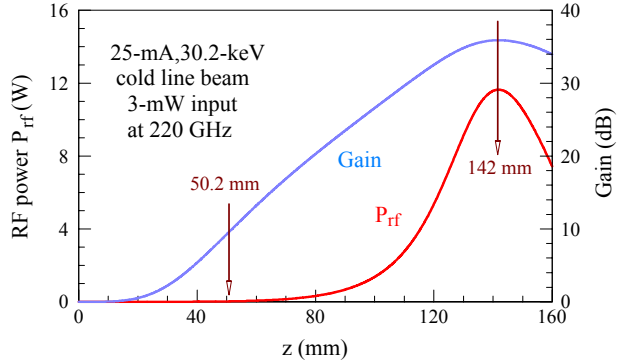

Fig. 5. Dependence of RF power $P_{rf}$ and gain on axial distance z for a rectangular dielectric-loaded TWT with a 25-mA, 30.2-keV cold line beam and a 3-mW input power of 220 GHz. The linear growth rate at z = 50.2 mm is 46.0/m, and the RF power reaches saturation (11.6 W) at 142 mm.

Figures 6 and 7 show the corresponding particle distributions and particle phase densities at different distances, illustrating how the beam is modulated and how the bunching center (density peak) evolves. It is seen from Figs. 6 and 7 that at z = 0 (entrance) the particle distribution is uniform and so is the density (50.8/rad), and the beam begins to get modulated by the RF field with the initial bunching center at the phase ($\omega t - k_z z$) = $\pi$ (radians). The density of the bunching center is increased to 58.3/rad when the center drifts to 1.8 (0.57$\pi$) at z = 50.2 mm, and some particles go into acceleration field from deceleration field. At z = 80 mm the bunching center is located at 1.4 (0.45$\pi$) with the density increased to 76.8/rad, and there are some particles falling into an adjacent RF period (from -2$\pi$ to 0) and beginning to form another bunching center. At z = 110 mm, two bunching centers are formed with a phase separation of ~2$\pi$ (= 1.70 + 4.57), and the density increased to ~320/rad. At this time, some particles around the first bunching center with the phase 1.70 (0.54$\pi$) have lost so much energy that their energy is lower than 30 keV, leading to ($\beta_0 - \beta_{ph}$) + $\delta\beta$ = ($\beta - \beta_{ph}$) < 0 [see Eq. (8)] and their phases begin to increase (instead of decease), as shown on the 142-mm distribution in Fig. 6, where the entire phase range is about 2.8$\pi$, compared with the injection phase range 2$\pi$. At saturation, as mentioned in the previous qualitative analysis, more and more particles begin to get out of deceleration fields into acceleration fields. This also can be seen from the 142-mm distribution where there is a significant part of particles between $\pi$ and 4 (acceleration phase).

It should be noted that the self-fields are ignored in the calculations. This effect is clearly observed in the 142-mm distribution in Fig. 6. For this distribution, the particle density may go to infinity at some phases because of no self-fields, and that is why its density curve is not shown in Fig. 7. However, if the axial self-field is included, later particles cannot catch up with or surpass the previous particles, and thus the phase of a particle entering the interaction region later is always larger than the ones of early particles at any given *z*-position; in other words, no two particles have the same phase,[25] and the particle phase density must be finite.

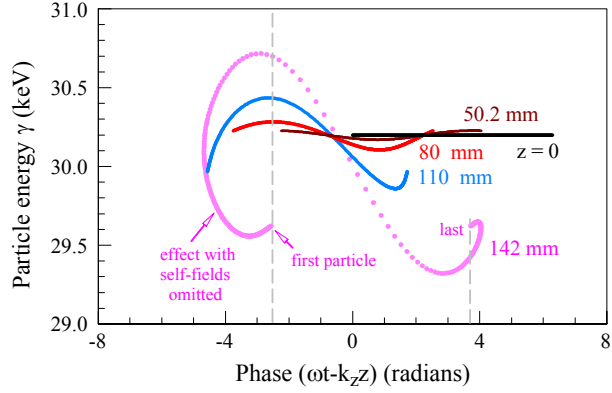

Fig. 6. Particle distributions on an energy-phase plane for axial distances $z = 0$ (initial), 50.2 mm (linear growth regime), 80 mm, 110 mm, and 142 mm (saturation). Note: for the 142-mm distribution there are observable effects resulting from ignoring self-fields in simulations: the part on the left side of the long vertical dash line, and the part on the right side of the short vertical dash line.

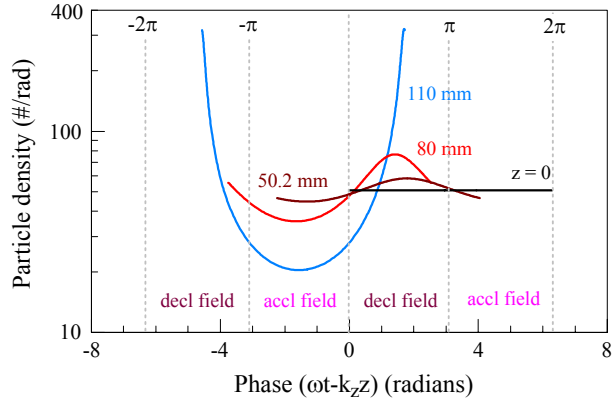

Fig. 7. Dependence of particle density on phase for axial distances $z = 0$, 50.2 mm, 80 mm, and 110 mm. It is seen that the density peak value gets larger as the peak drifts left, and two peaks with a phase separation of $2\pi$ are formed at $z = 110$ mm.

As we know, the beam energy has an important effect on synchronization with RF waves and the growth rate in the approximate linear regime depends on the beam energy; for example, 46.4/m for a 30.10-keV beam, 47.4/m for a 30.15-keV beam, and 46.0/m for a 30.20-keV beam. These growth rates are larger than but comparable with 40.2/m obtained from the linear theory in Sec. II.

Based on an ideal particle that gives up its maximum energy to the RF field at the end of the interaction region with a length of *L*, we can estimate the power conversion efficiency for a low efficiency device in such a way similar to the one used by Gold and Nusinovich,[30] given by

$$\frac{\Delta P_{rf}}{P_{b0}} \sim \gamma_0(\gamma_0 + 1)\frac{\lambda_g}{L} \approx \frac{2\lambda_g}{L}, \qquad \text{(for } \gamma_0 \approx 1\text{)} \qquad (9)$$

where $P_{b0}$ is the initial beam power, $\Delta P_{rf}$ is the gained RF power, and $\lambda_g$ is the waveguide wavelength. Eq. (9) indicates that the RF power conversion efficiency does not depend on the input RF power and beam current. Similarly, we also can estimate the required initial axial velocity (voltage) spread $\Delta\beta$ ($\Delta V$) that should be small enough to get this conversion efficiency, given by

$$\frac{\Delta\beta}{<\beta_0>} \approx \frac{1}{2}\frac{\Delta V}{<V_0>} < \frac{1}{2}\frac{\Delta P_{rf}}{P_{b0}} \sim \frac{\lambda_g}{L}, \qquad \text{(for } \gamma_0 \approx 1\text{)} \qquad (10)$$

where $\langle\beta_0\rangle$ and $\langle V_0\rangle$ are average initial particle velocity and voltage.

For the 30.2-kV beam case, taking $L = 142$ mm and $\lambda_g = 0.45$ mm, we have the power efficiency $\Delta P_{rf}/P_{b0}$ ~ $2\times 0.45/142 = 0.63\%$ estimated from Eq. (9), compared with $\Delta P_{rf}/P_{b0} = 11.6/755 = 1.54\%$ from the particle simulations shown in Fig. 5. The efficiency from Eq. (9) is considerably smaller, but comparable.

When circuit loss is considered, the beam-wave interaction will be weakened. Figure 8 shows the dependence of RF power on axial distance with a lossy circuit resulting from conducting walls (conductivity: $5.56\times 10^7$ siemens/m) and dielectric slabs (loss tangent: $5\times 10^{-4}$), compared with the one with a loss-free circuit in Fig. 5.

It is seen from Fig. 8 that the RF power is reduced at $z = 142$ mm because of the circuit loss, and in the initial part there occurs a power absorption, leading to a shrinking of flowing RF power. This shrinking comes from the fact that the RF power lost in the circuit is larger than the power that the electron beam gives up. Obviously this shrinking sets up a requirement on dielectric loss and puts a limit to the application of high-loss materials. As we have seen in Sec. II, the conventional Pierce three-wave theory also predicts a power absorption or initial RF power dip caused by its negative gain effect. And this dip has been observed experimentally by Caldwell.[31] From the above analysis, one might argue: since the circuit loss exists inevitably in a practical structure, probable it is not easy to judge whether the observed dip comes from three-wave interaction or from the circuit loss. This interesting topic is beyond the scope of the paper.

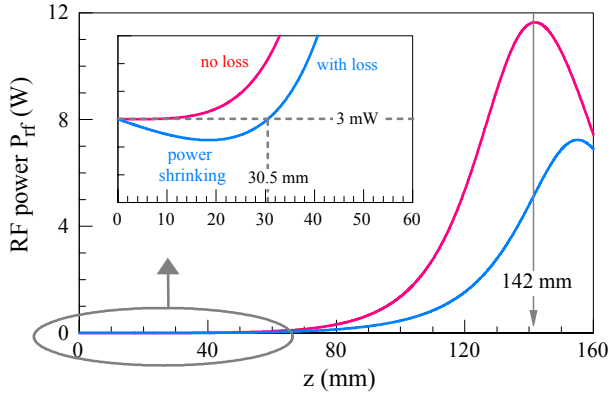

Fig. 8. Dependence of RF power $P_{rf}$ on axial distance $z$ with a lossy circuit, compared with the one with a loss-free circuit. The flowing RF power shrinks in the initial part because of the circuit loss.

## IV. INFLUENCES OF PARAMETERS ON TWT PERFORMANCE

In this section, influences of beam energy, RF frequency, velocity spread, dielectric loss, and input RF power on TWT performance are examined by simulations of a nominal 30-keV, 25-mA beam that has a finite cross section and finite axial velocity spread, and is focused by an applied uniform axial magnetic field.

The beam transverse dimensions are set to be 0.1mm×0.1mm, corresponding to a round solid beam with a radius of 56 microns, and the focusing magnetic field is 4 kG. Since the TWT interaction is based on longitudinal bunching and the beam is well focused, the initial transverse velocity spread is ignored to simplify calculations. The normalized axial velocity spread $\Delta\beta$ is taken to be $7.5\times 10^{-4}$, compared with the initial normalized thermal velocity spread ~ $(\kappa_B T/m_0 c^2)^{1/2} = 4.5\times 10^{-4}$ (corresponding to ~ 0.08-kV voltage spread for a 30-keV beam), where $\kappa_B$ is Boltzmann constant, and $T$ is taken to be 1200 degrees (K).[32,33] The particles are initially uniformly distributed, with 40 particles in one RF cycle from $\omega t_0 = 0$ to $2\pi$, with 9 particles in the 0.1mm×0.1mm beam cross section, and 3 particles for the axial velocity spread. Simulations show that the profile of flowing RF power obtained from one-cycle distribution is the same as the one from a multi-cycle distribution, since a steady-state RF flowing power is assumed. The dielectric loss tangent is taken to be $5\times 10^{-4}$ (silicon,[21] for example), and the conductivity of conducting walls is taken to be $5.56\times 10^7$ siemens/m.

### A. Influences of beam energy and RF frequency

In this sub-section, the influences of injection beam voltage and RF power frequency on beam-wave interaction are examined by (*i*). changing injection beam energy for a given RF frequency, (*ii*). changing RF frequency for a given beam energy, and (*iii*). changing beam energy to match RF frequency change to keep them in synchronization.

Figure 9 shows RF power profiles for different beam energies for a given 220-GHz, 3-mW input power. At $z$ = 142 mm, the RF powers are, respectively, 3.4 W for a 30.10-keV beam, 6.3 W for a 30.20-keV beam, 5.6 W for a 30.25-keV beam, and 2.6 W for a 30.28-keV beam. The circuit loss seems important; for example, the loss for the 30.2-keV case is 3.3 W, compared with its 6.3-W saturation power. It is seen that the initial power dip is a little increased with the decrease of the beam energy, because a higher energy beam has a larger equilibrium velocity and bunching centers shift into strong deceleration fields earlier, leading to the RF power getting gain earlier.

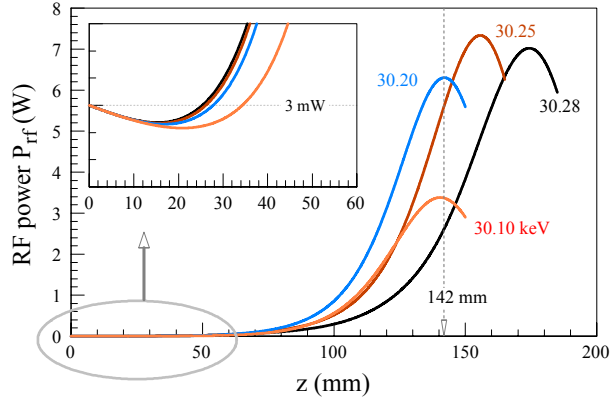

Fig. 9 (Given RF frequency). RF power profiles for different beam energies. Beam current: 25 mA; initial axial velocity spread: $7.5 \times 10^{-4}$ ($\Delta V/\langle V_0 \rangle$ = 0.46%); focusing magnetic field: 4 kG; input RF power: 3 mW at 220 GHz. For a 142-mm interaction length, the beam energy bandwidth is ~ 0.18 keV (0.6%) at FWHM, with a central beam energy of ~ 30.20 keV.

Figure 10 shows RF power profiles for different RF power frequencies for a given 30.20-keV beam. At $z$ = 142 mm, the RF powers are, respectively, 3.8 W for 219 GHz, 6.3 W for 220 GHz, 4.4 W for 220.7 GHz, and 1.6 W for 221 GHz.

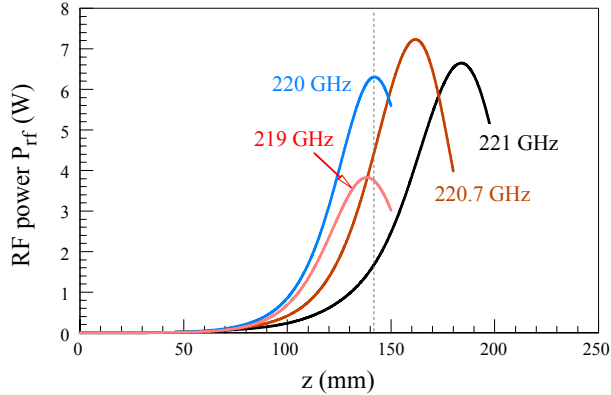

Fig. 10 (Given beam energy). RF power profiles for different RF frequencies for a 30.20-keV beam. It is seen that the frequency bandwidth is about 2 GHz (0.9%) at FWHM for a 142-mm interaction distance.

For a given RF frequency or beam energy, the corresponding energy or frequency bandwidth is small, both less than 1% from Figs. 9 and 10. This is because there is a required synchronous condition between the phase velocity and beam energy. For a given dispersive structure, the phase velocity varies with frequency, as shown in Fig. 2. If the injected beam energy can trace the change in frequency, then the limitation of bandwidth would mainly come from the RF coupler that feeds or extracts RF power. Figure 11 shows the power profiles for different frequencies for which synchronous injected beams are used. It is seen that the RF powers at $z$ = 142 mm are, respectively, 3.2 W for the 180-GHz, 35.60-keV case, 5.6 W for the 200-GHz, 32.45keV case, 6.3 W for the 220-GHz, 30.20-keV case, and 3.4 W for the 240-GHz, 28.70-keV case, and the bandwidth is up to 60 GHz. From this, we know that controlling beam energy to trace frequency change of input RF power can significantly improve the bandwidth, which might be interesting in engineering.

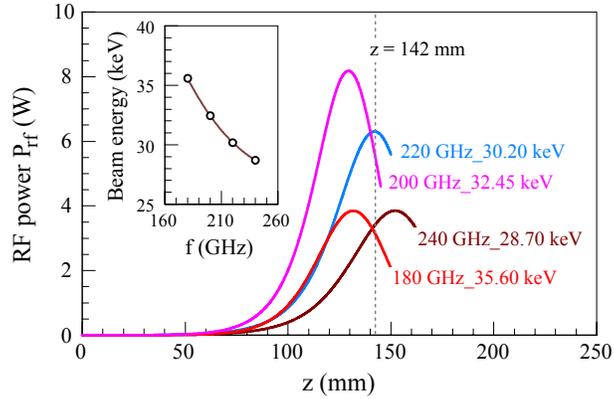

Fig. 11 (Injected beam energy matching RF frequency). RF power profiles for different beam energies and RF power frequencies. The frequency bandwidth is 60 GHz (~28%) at FWHM.

### B. Influence of beam velocity spread

Figure 12 shows the power profiles for different initial axial velocity spreads for a 30.2-kV, 25-mA beam, with a 3-mW, 220-GHz input power. It is seen that at $z$ = 142 mm the RF powers are, respectively, 6.3 W for a spread of $7.5\times10^{-4}$, 2.7 W for a spread of $1.5\times10^{-3}$, and 0.23 W for a spread of $2.0\times10^{-3}$. When the velocity spread is larger than $1.5\times10^{-3}$, corresponding to the relative spread $\Delta\beta/<\beta_0>$ = 0.46%, the power goes down much faster. The saturated RF power is also deceased when the velocity spread increases.

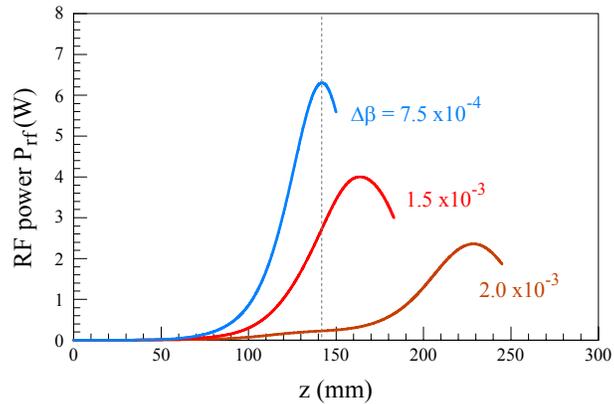

Fig. 12. RF power profiles for different initial axial velocity spreads for a 30.2-keV beam. RF power decreases with increase of velocity spread.

### C. Influence of dielectric loss

Figure 13 shows the power profiles for different dielectric loss tangent values for a 30.2-keV, 25-mA beam. It is seen that the saturated RF power is decreased when the loss tangent increases and the distance at saturation becomes longer; this is because a longer distance is needed to get the beam bunched due to the dielectric energy absorption. The saturated powers are, respectively, 6.3 W for a $5\times10^{-4}$ loss tangent, 3.6 W for a $2\times10^{-3}$ loss tangent, and 2.4 W for a $3\times10^{-3}$ loss tangent. Calculations show that if the loss tangent is larger than $5\times10^{-3}$, little power gain can be obtained.

### D. Influence of input RF power

Figure 14 shows the RF profiles for different input RF powers. It is seen that the saturated RF power varies little with input RF power. The input powers are, respectively, 3 mW, 6 mW, and 12 mW, while the saturated RF powers are only different by 3%, mainly resulting from circuit loss. This result is well consistent with Eq. (9), which predicts that the power conversion efficiency does not depend on input RF power for a low-efficiency device. This result is also consistent with the one given by Datta and coworkers for helix TWT 1D simulation.[34]

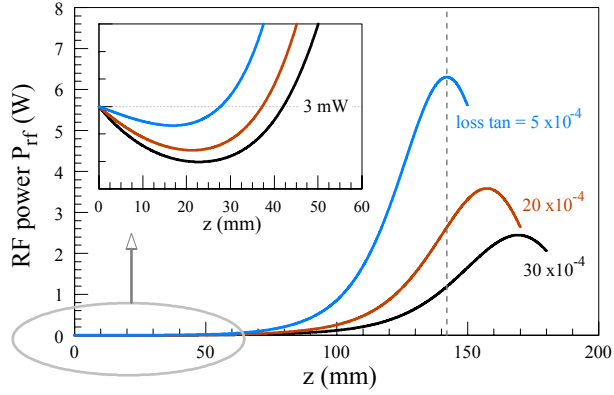

Fig. 13. RF power profiles for different dielectric losses. The RF powers at $z = 142$ mm are, respectively, 6.3 W, 2.6 W, and 1.2 W for three different loss tangent values.

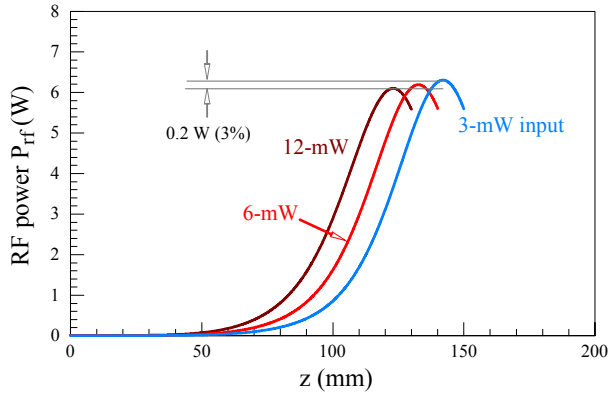

Fig. 14. RF power profiles for different input RF powers. The saturated RF powers are, respectively, 6.3 W at $z = 142$ mm for a 3-mW input, 6.2 W at $z = 133$ mm for a 6-mW input, and 6.1 W at $z = 123$ mm for a 12-mW input.

Since the interaction length at the saturated power reduces as with the increase of input power, the output power for a given interaction length will change when the input power is adjusted, as observed experimentally by Schächter and coworkers.[35]

## V. TRANSVERSE SPACE-CHARGE EFFECT

Space charge effect caused by self-fields is challenging. The self-electric field is dominant compared to the self-magnetic field for a low energy beam. Simulations for a 30.2-kV, 25-mA warm beam show that the RF power profiles for 1-kG and 8-kG focusing magnetic fields are almost the same. This is because the transverse space charge effect is ignored. For a well-focused round solid beam, it is assumed that the transverse space charge fields can be written in the form of [32]

$$E_r^{(sc)} = -\frac{I_0 Z_0}{2\pi r_b \beta_0} \frac{r}{r_b}, \qquad (r < r_b) \tag{11}$$

$$E_r^{(sc)} = -\frac{I_0 Z_0}{2\pi r_b \beta_0} \frac{r_b}{r}, \qquad (r > r_b) \tag{12}$$

where $Z_0 = 120\pi$ Ohms is the vacuum wave impedance, $I_0$ is the beam current, and $r_b$ is the beam radius. For $I_0 = 25$ mA, $r_b = 0.056$ mm, and $\beta_0 = 0.3284$ (30 keV), the field at the beam edge is ~ 82 kV/m, although the depressed voltage across the beam cross section is less than 2.3 volts. This field strength is much larger than the maximum field (38.6 kV/m) in the structure created by a 1-W flowing RF power (see Fig. 1). Simulations indicates that this space-charge field (central force field) combined with the focusing magnetic field causes electrons to make Larmor precession, bringing about an additional velocity spread by absorbing input RF power, possibly leading to stalling the bunching-buildup process. To prevent this, one may increase the focusing magnetic field to reduce beam energy absorption, or increase the input RF power to compensate the part of the power to be absorbed by the beam. To better understand the physics, simulations have been done and they are presented below.

In computations, the same beam as the one in Fig. 10 is employed, while the transverse space-charge effect is considered through Eqs. (11) and (12). Figure 15 shows the RF power profiles for 8-kG and 20-kG focusing magnetic fields. It is seen in the power absorption region that the RF power ripples because of particles' precession and the rippling amplitude for the 20-kG magnetic field is much smaller. The ripples mean that the beam and the RF wave exchange energy by convection. This energy convection may stop beam bunching in the initial bunching buildup process, when the focusing field or input power is not strong enough.

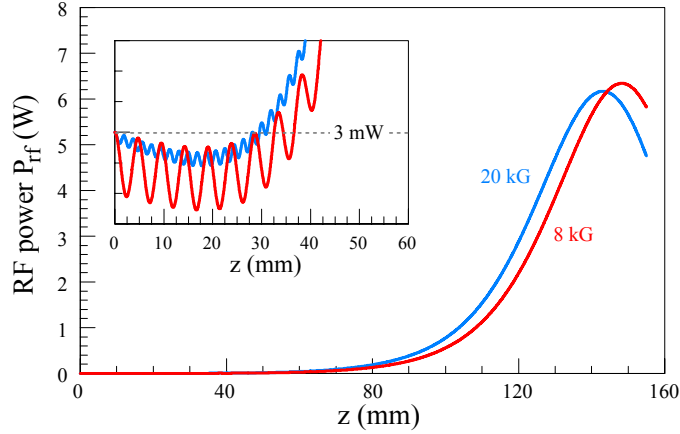

Fig. 15. Dependence of RF power $P_{rf}$ on axial distance $z$ for 8-kG and 20-kG focusing magnetic fields. The input RF power is 3 mW at 220 GHz. The injected 30.2-keV, 25-mA beam has an axial velocity spread of $\Delta\beta = 7.5\times10^{-4}$. The RF power fluctuations for the beam focused by a 20-kG magnetic field are much smaller in the initial low power region.

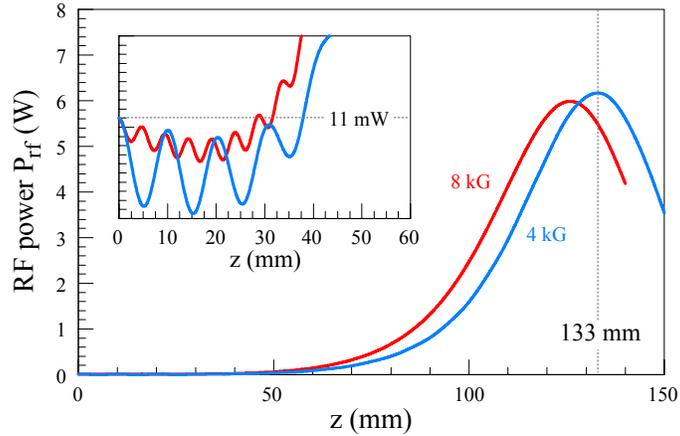

Fig. 16. Dependence of RF power $P_{rf}$ on axial distance $z$ for 4-kG and 8-kG focusing magnetic fields. The input RF power is increased to 11 mW while the injected beam is the same as in Fig. 15. The RF power for the beam focused by a 4-kG magnetic field fluctuates more deeply in the initial low power region.

To reduce focusing magnetic field, the input RF power has to be increased to avoid the bunching stalled, although resulting in a reduce in power gain. As shown in Fig. 16, the input power for the two cases is increased to 11-mW, while the focusing magnetic fields are reduced to 4 kG and 8 kG respectively. It is seen that the saturated powers are very close to the ones in Fig. 15, just with shortened interaction distances due to the increased input power as indicated in Fig. 14. The trajectory for a sample electron initially at the beam edge is shown in Fig. 17. The beam channel size is 0.2mm×1.2mm (Conf. Fig. 1), and all electrons pass through the channel without hitting walls.

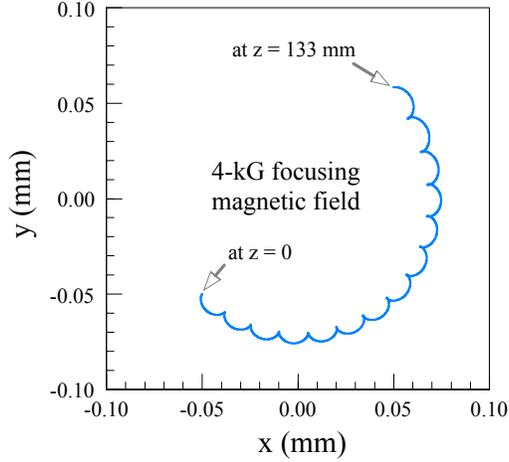

Fig. 17. Orbit projected on *x-y* plane (0 ≤ *z* ≤ 133 mm) for a sample electron at the beam edge. Note that the electron gyrates anti-clockwise around the focusing magnetic field $\mathbf{B}_0$ (looking towards the page), while the guiding center drifts along the $\mathbf{E}^{(sc)} \times \mathbf{B}_0$ direction.

## VI. SIMULATION DESIGN OF RF COUPLERS

Input/output RF couplers for dielectric-loaded TWTs are indispensable elements and bandwidth of the couplers is an important issue. RF wave reflections are often provoked in non-uniform regions. Using reflections to reduce or cancel reflections is often adopted in designs of RF transmission systems; slowly changing structure parameters helps in increase of bandwidth, while fast changing structure parameters tends to result in decrease of bandwidth. In this section, a proof-of-principle RF simulation design of couplers is presented.

It is seen from Fig. 1(b) that the *E*-fields of the $LSM_{21}$ operating mode are mirror-symmetric with respect to the *y-z* plane and the *x-z* plane (not shown), and $E_x$ reaches the maximum on the two side walls. From this, a coupler with two ports having mirror-symmetric *E*-polarizations in the *x*-direction may meet the above two symmetries. Obviously, the wave propagation direction in such a coupler should be the same as in the interaction region. To reduce reflection, the loaded dielectric slabs and conducting walls should be tapered. Based on these guidelines, a suggested coupler scheme is shown in Fig. 18 where a shortened version is taken to save computation time. Reducing the slope of tapers may reduce reflections but the couplers get longer. Such a coupling scheme has been successfully used in a cold design of 10-GHz vane-loaded rectangular waveguide structure to make a proof-of-principle for a 300-GHz TWT structure.[36,37]

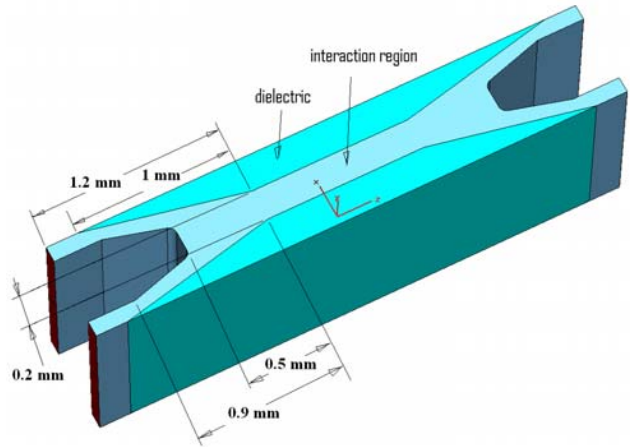

FIG. 18. A shortened 220-GHz TWT structure with input/output couplers in simulations, in which there are one 1-mm interaction region loaded with uniform dielectric slabs and two 1-mm regions loaded with tapered slabs, and the waveguide walls are also tapered following the slabs. All the parameters in the uniform region are the same as in Table I. The ports have transverse dimensions of 0.1mm×1.2mm and the total structure length is 3.4 mm.

In computations, the dielectric loss tangent is taken to be $5\times10^{-4}$ but the conducting wall loss is ignored. RF simulations show that a dominant traveling wave is excited at design $LSM_{21}$ mode with an 8-GHz bandwidth, within which the power transmission coefficient is above ~ 90%.

The S-parameters obtained from simulations are shown in Fig. 19, and only four of the 16 S-parameters are different due to the structure's symmetry. It is seen that the voltage transmission coefficient $s_{31}$ from port-1 to port-3 has a very small change within an 8-GHz bandwidth. The power transmission coefficient from port-1 to port-3 and port-4 is given by $s_{31}^2+s_{41}^2$, corresponding to 93% at 220 GHz. From energy conservation law, it can be estimated that the power lost in the 3-mm long dielectric slabs (1-mm uniform region and 2-mm tapers) is $1-(s_{11}^2+s_{21}^2+s_{31}^2+s_{41}^2)$ = 2.1%, or 0.7%/mm, corresponding to an attenuation constant of 3.5/m, much less than the growth rate 40.2/m from the linear theory given in Sec. II.

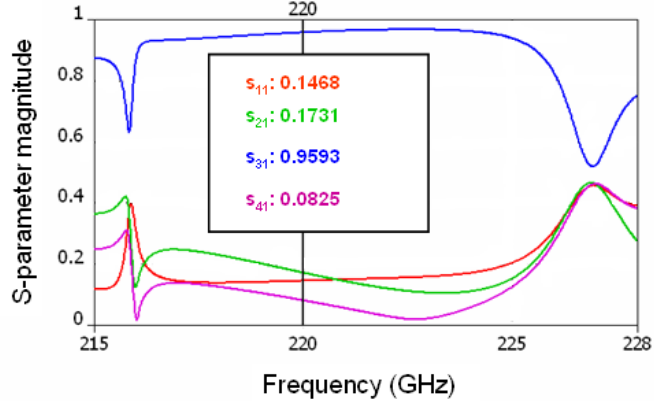

Fig. 19. S-parameters *vs* frequency for a 220-GHz TWT structure with input/output couplers. The power transmission coefficient is between 89% and 94% in a bandwidth of 8 GHz (217 GHz – 225 GHz), and it is ~ 93% at 220 GHz with a voltage standing wave ratio (VSWR) of 1.34.

The structure configuration shown in Fig. 18 does not include input/output beam channels for the sake of simplicity. RF simulations show that the S-parameters have little change when a 0.2 mm×0.4 mm beam channel is added, because the RF waves are prevented from going into the channel for its small size. In such a case, the power transmission coefficient varies from 88% to 94% within the same 8-GHz bandwidth.

Figure 20 shows the contour of the amplitude of complex electric field (= Max |**E**| over time) on the symmetric z-x plane. As we know, if the amplitude of a wave is constant along the propagation direction in a regular waveguide system, then the wave is a pure traveling wave. It can be seen that the field amplitude in the uniform region varies along the z-direction because there is a small standing-wave component. The fields around the inner metal corners of the input ports 1 and 2 are much stronger, but below 200 kV/m. Analytical theory indicates that on the conducting walls for an infinite structure, the maximum electric field $E_{xmax}$ is about 55 kV/m for a pure traveling-wave produced by a 2-W flowing RF power. By checking the color scale for a rough estimation, it is found that the peak- and node-field values on the walls in the uniform interaction region are, respectively, in the range of 60-70 kV/m (~ $1+s_{11}$) and 40-50 kV/m (~ $1-s_{11}$), and they are in reasonable agreement with 55 kV/m if considering that the voltage reflection coefficient $s_{11}$ is about 0.15 from Fig. 19.

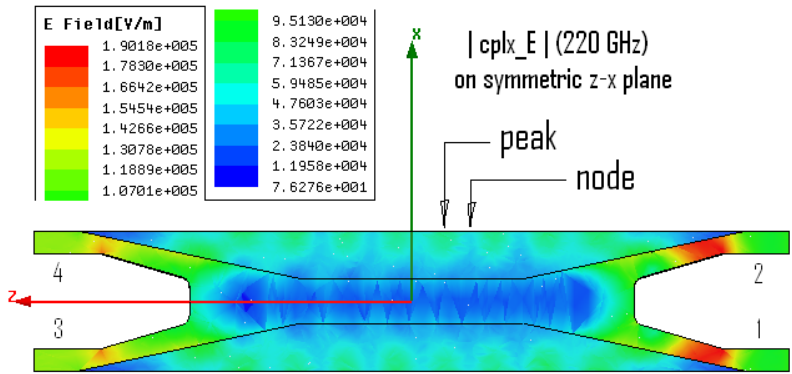

Fig. 20. Contour of amplitude of complex electric field on symmetric z-x plane. The fields are excited at port-1 and port-2 simultaneously with 1 W/port.

Mode identification is an important part in RF simulations. Identification of *y*-mode index is easy for this structure because the dielectric is uniform along the *y*-direction. Identification of mode in the *z*- and *x*-directions is based on two points: waveguide wavelength at 220 GHz and the symmetry of fields on the symmetric *z-x* plane. The waveguide wavelength can be estimated by counting wave peaks in the 1-mm uniform interaction region. It is seen from Fig. 21 that there are about 4.5 peaks, well corresponding to the theoretical waveguide wavelength 0.45 mm. In addition, we find that $E_x$-fields on the two side conducting walls are mirror-symmetric with respect to the *y-z* plane, consistent with the one from the analytic theory shown in Fig. 1(b).

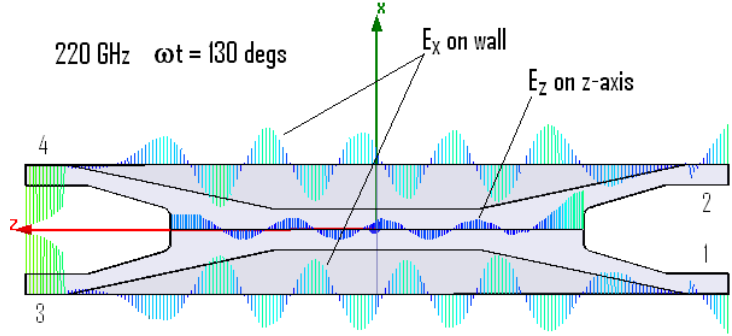

Fig. 21. Electric field **E**-dependence on central lines on side conducting walls ($x = \pm 0.31$ mm and $y = 0$) and on *z*-axis ($x = 0$ and $y = 0$). On the conducting walls, $\mathbf{E} = \mathbf{E}_x$; on the *z*-axis, $\mathbf{E} = \mathbf{E}_z$. It is seen that there is a $\pi/2$-phase difference between $E_x$ and $E_z$; the interaction region contains about 4.5 wave peaks, and the ratio of the height of $E_x$-amplitude to the height of $E_z$-amplitude is about 4, compared with 4.2 from analytic theory.

It is seen from Fig. 21 that the electric field on the dielectric taper end has a big jump. This can be understood from the boundary conditions on the common intersection line of vacuum, dielectric, and perfect conducting wall, given by

$$(\varepsilon_r - 1) E \sin \alpha = 0, \quad (13)$$

$$(\mu_r - 1)\left( H^2 - \left[ \mathbf{H} \cdot \frac{(\mathbf{n}_d \times \mathbf{n}_c)}{|\mathbf{n}_d \times \mathbf{n}_c|} \right]^2 \right)^{1/2} \cos \alpha = 0, \quad (14)$$

where $\mathbf{n}_d$ and $\mathbf{n}_c$ are, respectively, the unit normal vectors of the dielectric and conductor planes, as shown in Fig. 22. It is found from Eq. (13) that $E = 0$ must hold on the intersection line if $\varepsilon_r \neq 1$ and $\sin \alpha \neq 0$, which explains the jump.

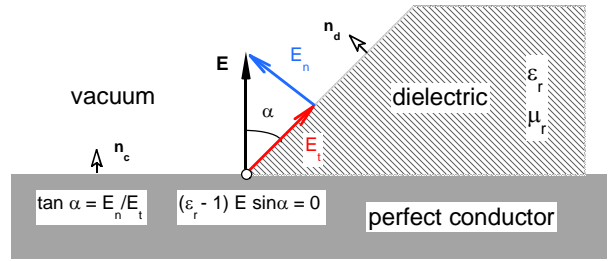

Fig. 22. Illustration of boundary conditions on the common intersection line of vacuum, dielectric, and perfect conducting wall. On the conducting wall, the electric field only has the normal component while the magnetic field may have two tangent components (not shown).

## VII. CONCLUSIONS AND REMARKS

In this paper, simulation results for a 220-GHz dielectric-loaded traveling-wave amplifier have been presented. The RF gain profiles from simulations and Pierce three-wave theory are in reasonable agreement. An intuitive way to show the evolution of beam bunching centers during beam-wave interaction is provided. For a nominal 30-keV, 25-mA beam, about 6-W output power can be obtained with a bandwidth of about 2 GHz (0.9%) at FWHM for a given beam energy. In some practical applications, a wideband frequency-adjustable source may be desirable, and setting beam voltage adjustability to match frequency sweeping might be an option, for which simulations have shown a well improved bandwidth (up to 60 GHz, ~ 28%). Dielectric loss has an important effect on beam-wave interaction and the loss tangent should be less than $2\times10^{-3}$ to make the device have enough output. A focusing magnetic field more than 4 kG, depending on required RF power gain, is needed for beam bunching buildup without interruption by transverse space-charge effect.

This amplifier seems to have a severe requirement on beam axial velocity spread, and a relative spread of less than 0.5% is needed. This spread is just a few times of the thermal velocity spread produced by thermionic cathodes.[32,33] As the development of electron-source technology, however, beam quality and feature will be further improved; for example, the carbon nanotube cold cathode might be one of promising candidates, which can directly provide a prebunched beam to greatly enhance the beam-wave interaction.[11]

A scheme of symmetric excitation for RF couplers has been presented. The significance of the scheme is improving mode suppression and bandwidth, based on the fact that a symmetric excitation naturally kills all non-symmetric modes. Theoretic and simulation analyses indicate that anti-symmetric $LSM_{11}$ mode is a strongly competing mode. If a non-symmetric excitation is employed (one-port coupler, for example), the $LSM_{11}$ mode may be excited in preference. In principle, the $LSM_{11}$ mode can be suppressed by putting slots on proper symmetric planes. But practical slots must have a finite width that may affect the design $LSM_{21}$ mode, and possibly creates new interfering modes. Moreover, a non-symmetric excitation has already changed the geometric and electromagnetic symmetry, and the proper locations for slots may not be easy to control. More importantly, the symmetric excitation will provide a symmetric RF field ($E_z$) in the interaction region, greatly enhancing the beam-wave energy exchange. Of course, nothing is free. A symmetric excitation coupler needs additional microwave elements; for example, power divider for input signal, and waveguide combiner for output RF power.

In summary, relatively compact, moderate low-energy beam TWT amplifiers for THz-source applications are possible in principle although there are some challenges in technology, such as generation of low spread beams, control of tolerance in micromachining dielectric slabs, and calibration of the structure and focusing magnetic field.

## ACKNOWLEDGMENTS

The author would like to thank Prof. J. L. Hirshfield for his helpful discussions and critical comments. It is impossible to result in the present work without his support when the author was in Omega-P, Inc.

## APPENDIX: IMPROVEMENT OF NUMERICAL METHOD

Numerical errors in computations are inevitable. Some errors are easy to judge while some others may not, possibly misleading to incorrect physical conclusions. In this appendix, an example for such numerical errors in RF power computations is given and a method of how to improve the computations is suggested.

Theoretical and numerical analysis indicates that when taking an average over particle energy during the initial beam bunching buildup, a commonly-used direct method may result in misleading numerical errors and a physical explanation based on the errors is the negative gain effect, i.e., EM power is transferred to the beam.[16,20] Assigning this explanation is probably prompted by Pierce three-wave theory which does predict such a phenomenon, depending on circuit parameters.[31] The cause of this problem comes from the fact that the energy exchange is over-weak and the average particle energy converges slowly with the particle number, leading to the method not performing well on precise computations. .

Particle dynamics calculations for a cold line beam come down to solving the classic linear accelerator equation

$$\frac{d\gamma}{dz} = \frac{eE_0}{m_0 c^2}\sin\psi, \quad \text{with} \quad \psi = \omega t - k_z z. \tag{A1}$$

To simplify the analysis, we assume the RF electric amplitude $E_0$ is constant, which is a good approximation in the initial beam bunching process. This equation has an exact analytical solution with respect to $\psi$;[38] however, a solution with respect to the independent variable $z$ is not easy to obtain. If the beam initial velocity is not equal to the operating mode phase velocity ($\beta_0 \neq \beta_{ph}$), and the perturbation parameter $|\lambda_0 eE_0/m_0c^2|\ll 1$ holds, Eq. (A1) can be linearized. From this, we can obtain a linear perturbation solution for normalized average energy increment, given by

$$<\gamma - \gamma_0> = \frac{2\pi}{\beta_0^3}\left(\frac{\lambda_0 eE_0}{m_0 c^2}\right)^2 \frac{1}{\hat{k}_z^3 \lambda_0^3}\left(\cos(\hat{k}_z z) - 1 + \frac{1}{2}(\hat{k}_z z)\sin(\hat{k}_z z)\right), \tag{A2}$$

where $<\gamma-\gamma_0>$ means taking the average over the initial phase $\omega t_0$ for all particles uniformly distributed from $\omega t_0 = 0$ to $2\pi$, $k_{\bar{z}} = k_z(\beta_0 - \beta_{ph})/\beta_0$, and $\lambda_0$ is the wavelength in free space. It should be noted that the linear electron efficiency $<\gamma_0-\gamma>/(\gamma_0-1)$ for TWT amplifiers can be directly obtained from Eq. (A2), and the form of the expression obtained is exactly the same as the one for coherent synchrotron radiators.[6,7]

If $z/\lambda_0 << 1$ holds, Eq. (A2) can be further simplified, to obtain

$$<\gamma-\gamma_0> \approx -\frac{\pi^2}{6}\left(\frac{\lambda_0 e E_0}{m_0 c^2}\right)^2 \frac{(\beta_0 - \beta_{ph})}{\beta_0^4 \lambda_0^3 \lambda_g} z^4 . \tag{A3}$$

From above, we arrive at a strict qualitative theoretical result: at the very beginning ($z/\lambda_0<<1$), $<\gamma-\gamma_0> \leq 0$ if $\beta_0 > \beta_{ph}$, while $<\gamma-\gamma_0> \geq 0$ if $\beta_0 < \beta_{ph}$; in other words, the beam loses energy (positive RF power gain in TWTs) if the beam initial velocity is larger than the phase velocity, while the beam gains energy (negative RF power gain in TWTs) if the beam initial velocity is less than the phase velocity.

So far we have obtained necessary theoretical results. Now let us examine numerical methods, results, and errors. The direct method for taking an average is defined by

$$<\gamma-\gamma_0> = \frac{1}{N}\sum_{i=1}^{N}\left[\gamma(\omega t_0^{(i)}, z) - \gamma_0\right]. \tag{A4}$$

This is a straightforward way to evaluate an average. To improve the precision, a suggested method is given by

$$<\gamma-\gamma_0> = \frac{1}{2\pi}\int_0^{2\pi}\left[\gamma(\omega t_0, z) - \gamma_0\right] d(\omega t_0), \tag{A5}$$

where cubic spline interpolation is used to construct the integrand in computations by calling routine DCSINT, for example, from Fortran IMSL math library, and then call routine DCSITG to carry out the integration. Let us call this technique a cubic spline integration (cs_itg) method for the convenience of description.

To carry out the numerical analysis, both the direct method and the cs_itg method are employed to calculate the energy increment $<\gamma-\gamma_0>$ in computing Eq. (A1). The beam energy is taken to be 30.2 keV ($\beta_0 = 0.3294$), the field amplitude $E_0$ is taken to be 0.505 kV/m corresponding to a 3-mW RF flowing power for the LSM$_{21}$ mode operating at 220 GHz, with $\lambda_0 = 1.3627$ mm, $\lambda_g = 0.4475$ mm, and $\beta_{ph} = 0.3284$ (30 keV measured in electron energy). The data given above are complete for solving Eq. (A1).

Figure 23 shows the dependence of energy increment $<\gamma-\gamma_0>$ measured in eV on axial distance $z$ for different particle numbers for the direct method. This method creates considerable errors. Obviously, the results in Fig. 23 are not consistent with Eq. (A3) which predicts $<\gamma-\gamma_0> \leq 0$ at the very beginning for $\beta_0 > \beta_{ph}$. It is seen that for the 64- and 320-particle cases, the maximum beam energy gains would be, respectively, 0.065 eV and 0.003 eV, corresponding to 1.63-mW and 0.07-mW RF power absorption out of the 3-mW RF input for a 25-mA beam, that is, 3.4-dB and 0.1-dB power attenuations. Calculations show that for a 32-particle case the beam energy gain would be 0.23 eV, corresponding to a 5.7-mW absorption power, even larger than the input RF power.

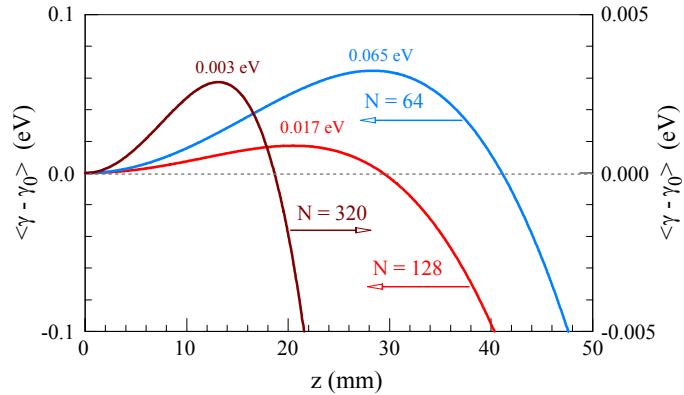

Fig. 23 (direct method). Average energy increment $<\gamma-\gamma_0>$ vs axial distance $z$ for different particle numbers. It is seen that $<\gamma-\gamma_0> >0$ holds for small $z$ ($\neq 0$), which contradicts the strict qualitative result Eq. (A3).

Figure 24 shows the dependence of energy increment $<\gamma-\gamma_0>$ on axial distance $z$ for different particle numbers for the cs_itg method. To compare, the linear perturbation solution from Eq. (A2) is also given. It is seen that the result for the 32-particle case is the same as the one for the 320-particle case, and they are completely consistent with the theoretical prediction of Eq. (A3). Surprisingly, the perturbation solution is so closely following the cs_itg solution over a 160-mm interaction distance, probably because the perturbation parameter $|\lambda_0 e E_0/m_0 c^2| = 1.3\times 10^{-6}$ is so small that the result converges fast.

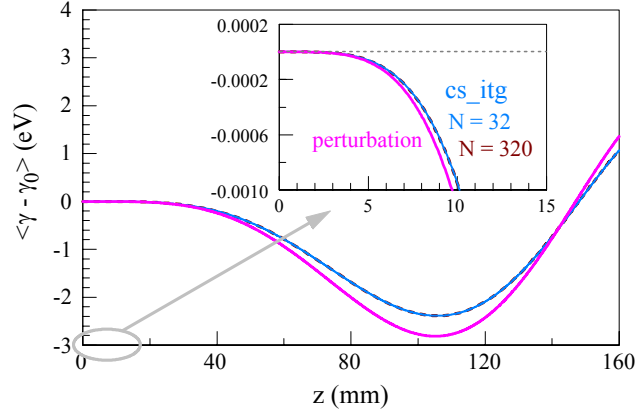

Fig. 24 (cs_itg method). Average energy increment $<\gamma-\gamma_0>$ vs axial distance $z$ for different particle numbers, with linear perturbation solution shown for comparison. It is seen that the curve for the 32-electron case exactly overlaps the one for the 320-electron case. The perturbation solution closely follows the curve obtained from the cs_itg method over a 160-mm distance.

From the above numerical analysis, we find that the cs_itg method has a much better precision; for example, even the precision of the 32-particle case for the cs_itg method is better than the precision of the 320-particle case for the direct method. We find that the results of the perturbation theory and cs_itg method are very consistent, and we may conclude that the power gain is positive during the initial beam bunching buildup process when $\beta_0 > \beta_{ph}$. However if the direct method is used in particle simulations, the numerical errors may lead to a negative gain, which is not real physics. The above conclusions have been confirmed by simulations for a TWT amplifier injected with a finite transverse cross section beam but without any circuit loss.

Although Eq. (A1) is an idealized dynamic equation for TWT interaction, it gives a conclusion that the direct method for taking an average does cause considerably larger numerical errors. In various non-linear theories, all calculations of RF flowing powers involve taking an average. Therefore, the conclusion has a general significance.

[1] M. Tonouchi, *Nature Photonics* **1**, 97 (2007).
[2] P. H. Siegel, *IEEE Trans. Microwave Theory Tech.* **50,** 910–928 (2002).
[3] P. G. O'Shea and H. P. Freund, *Science* **292**, 1853 (2001).
[4] G. L. Carr, M. C. Martin, W. R. McKinney, K. Jordan, G. R. Neil, and G. P. Williams, *Nature* **420**, 153 (2002).
[5] F. Wang, D. Cheever, M. Farkhondeh, W. Franklin, E. Ihloff, J. van der Laan, B. McAllister, R. Milner, C. Tschalaer, D. Wang, D. F. Wang, A. Zolfaghari, T. Zwart, G. L. Carr, B. Podobedov, and F. Sannibale, *Phys. Rev. Lett.* **96**, 064801 (2006).
[6] Changbiao Wang, *Phys. Rev.* **A 38**, 6215 (1988).
[7] Changbiao Wang, *Appl. Phys. Lett.* **53**, 1911 (1988).
[8] D. W. Sprehn, C. L. Rettig, and N. C. Luhmann, Jr., *Rev. Sci. Instrum.* **63**, 485 (1992).
[9] S. E. Korbly, A. S. Kesar, J.R. Sirigiri, and R. J. Temkin, *Phys. Rev. Lett.* **94**, 054803 (2005).
[10] M. K. Hornstein, V. S. Bajaj, R. G. Griffin, and R. J. Temkin, *IEEE Trans. Plasma Sci.* **35**, 27 (2007).
[11] K. B. K. Teo, E. Minoux, L. Hudanski, F. Peauger, J. Schnell, L. Gangloff, P. Legagneux, D. Dieumegard, G. A. J. Amaratunga, and W. I. Milne, *Nature* **437**, 968 (2005).
[12] P. H. Siege, A. Fung, H. Manohara, J. Xu, and B. Chang, "Nanoklystron: A monolithic tube approach to terahertz power generation," in *12th Int. Space Terahertz Technol. Symp.*, San Diego, CA, Feb. 2001, pp. 81–90.
[13] K. L. Felch, K. O. Busby, R. W. Layman, D. Kapilow, and J. E. Walsh, *Appl. Phys. Lett.* **38**, 601 (1981).
[14] H. P. Freund and A. K. Ganguly, *Phys. Fluids B* **2**, 2506 (1990)

[15] J. Gao and F. Shen, *Phys. Rev.* **A 73**, 043801 (2006).

[16] L. Schächter, *Phys. Rev.* **A 43**, 3785 (1991).

[17] G. Ling and Y. Liu, *Phys. Rev.* **E 50**, 4262 (1994).

[18] Changbiao Wang and J. L. Hirshfield, *Phys. Rev. ST Accl. and Beam* **9**, 031301 (2006).

[19] Changbiao Wang and J. L. Hirshfield, *Phys. Rev.* **E 51**, 2456 (1995).

[20] Y. Yang and W. Ding, *Phys. Rev.* **E 61**, 4450 (2000).

[21] M. N. Afsar and K. J. Button, *Proc. IEEE* **73**, 131 (1985).

[22] J. E. Walsh, T. C. Marshall, M. R. Mross, and S. P. Schlesinger, *IEEE Trans. MTT*-**25**, 561 (1977). One may assume $\omega$ is complex while $k_z$ is real in seeking usual traveling-wave solutions as shown in this reference, and the growth rate obtained is related to the cubic roots of (+1), which is consistent with energy conservation law because there is an opposite sign in front of $\omega$ and $k_z$ in $\exp[i(\omega t - k_z z)]$. But for traveling-wave solutions the physical explanation is not so self-consistent as the one based on the assumption of real $\omega$ and complex $k_z$.

[23] H. L. Andrews and C. A. Brau, *J. Appl. Phys.* **101**, 104904 (2007).

[24] L. Schächter, *Beam-Wave Interaction in Periodic and Quasi-Periodic Structures* (Springer-Verlag, Heidleberg, 1997), Chap. 4.

[25] A. S. Gilmour, *Principle of traveling wave tubes* (Artech House, MA, 1994), Ch. 10, p. 273; Ch. 9, p. 239.

[26] J. R. Pierce, *Traveling-Wave Tubes* (D. Van Nostrand Company, Inc., New York, 1950), Ch. IX, p. 134.

[27] W. Gai, P. Schoessow, B. Cole, R. Konecny, J. Norem, J. Rosenzweig, and J. Simpson, *Phys. Rev. Lett.* **61**, 2756 (1988).

[28] L. Schächter and J. A. Nation, *Phys. Rev.* **A 45**, 8820 (1992).

[29] K. R. Chu, *Review of Modern Physics* **76**, 489 (2004).

[30] H. Gold and G. S. Nusinovich, *Rev. Sci. Instrum.* **68**, 3945 (1997).

[31] G. R. Brewer and C. K. Birdsall, *IRE Trans. Electron devices*, 140 (1957).

[32] S. Humphries, *Charged Particle Beams* (John Wiley and Sons, NM, 1990), Ch. 7, p.285; Ch. 5, p. 206.

[33] W. B. Herrmannsfeldt, *SLAC-PUB*-**4843**, 1989, p. 3.

[34] S. K. Datta, S. U. M. Reddy, B. N. Basu, and K. U. Limaye, *Microwave and Optical Technology Letters* **18**, 310 (1998).

[35] L. Schächter, J. A. Nation, and D. A. Shiffler, *J. Appl. Phys.* **70**, 114 (1991).

[36] F. L. Krawczyk, B. E. Carlsten, L. M. Earey, F. E. Sigler, and M. E. Schulze, "Design of a 300 GHz broadband TWT coupler and RF-structure," in *Proc. of LINAC 2004*, THP84, p. 794.

[37] B. E. Carlsten, *Physics of Plasmas* **12**, 5088 (2002).

[38] J. R. Terrall and J. C. Slater, *J. Appl. Phys.* **23**, 66 (1952).